\documentclass[twocolumn,aps,pra,showpacs,raggedbottom,nobalancelastpage,amssymb,groupedaddress]{revtex4-1}

\usepackage{graphicx}
\usepackage{amsmath}
\usepackage{amssymb}
\usepackage{bm}
\usepackage[usenames]{color}
\usepackage{amsfonts}
\usepackage{appendix}
\usepackage{dsfont}

\def \Fst{$1^\textrm{st}$ }
\def \Snd{$2^\textrm{nd}$ }

\definecolor{Nathanblue}{rgb}{0.,0.24,0.51}

\def\be{\begin{equation}}
\def\ee{\end{equation}}

\begin{document}
\title{Four-dimensional topological lattices through connectivity}
\author{Hannah M. Price}
\affiliation{\mbox{School of Physics and Astronomy, University of Birmingham, Edgbaston, Birmingham B15 2TT, United Kingdom}}

%\date{\today}

\begin{abstract} 
Thanks to recent advances, the 4D quantum Hall (QH) effect is becoming experimentally accessible in various engineered set-ups. In this paper, we propose a new type of 4D topological system that, unlike other 2D and 4D QH models, does not require complicated (artificial) gauge fields and/or time-reversal symmetry breaking. Instead, we show that there are 4D QH systems that can be engineered for spinless particles by designing the lattice connectivity with real-valued hopping amplitudes, and we explain how this physics can be intuitively understood in analogy with the 2D Haldane model. We illustrate our discussion with a specific 4D lattice proposal, inspired by the widely-studied 2D honeycomb and brickwall lattice geometries. This also provides a minimal model for a topological system in Class AI, which supports nontrivial topological band invariants only in four spatial dimensions or higher. 
\end{abstract}

%\pacs{67.85.-d, 03.65.Sq, 37.10.Jk, 73.43.-f}

\maketitle

\section{Introduction} 

Topological concepts provide a powerful way to discover and classify different phases of matter~\cite{RMP_TI, RMP_TI2,bernevig2013topological}. Within this paradigm, topological states are characterised by topological invariants and classified according to symmetries and spatial dimensionality~\cite{Kitaev:2009AIP, Ryu:2010NJP,Chiu:2016RMP,Kruthoff:2017}.  As topological invariants are integer-valued, these states can exhibit remarkably robust phenomena, such as quantized transport and surface modes, unaffected by small perturbations. 

A notable example of a topological phase of matter is a 2D quantum Hall (QH) system~\cite{Klitzing:1980PRL}, which has energy bands indexed by the \Fst Chern number (1CN). The 1CN is a 2D topological invariant that is only non-zero if time-reversal symmetry (TRS) is broken, and which underlies the precise quantization of Hall conductance in the 2DQH effect~\cite{TKNN}. Although this effect was first discovered for electrons in a magnetic field~\cite{Klitzing:1980PRL}, such 2DQH states have recently been realised for different systems, such as cold atoms~\cite{Goldman:2016NatPhys,cooper2018} and photons~\cite{lu2016topological, khanikaev2017two, ozawareview}, thanks to the development of artificial gauge fields. 

The engineering of topology for artificial systems has also revitalised interest in topological physics predicted in higher spatial dimensions. This includes the 4DQH effect, which is a quantized current response, like its 2D cousin, but now due to a 4D topological invariant, called the \Snd Chern number (2CN)~\cite{Nakahara, Avron1988,frohlich2000,IQHE4D,QiZhang,edge2012metallic,Sugawa:2016arXiv}. Recently, this 4DQH effect has been probed using 2D ``topological pumps" of atoms~\cite{ Lohse2018} and photons~\cite{Zilberberg2018}, exploiting a mapping from higher-dimensional topological systems to lower-dimensional time-dependent pumps~\cite{Thouless83, Kunz1986, Kraus:2012a, Kraus:2012b, Kraus2013, Verbin:2015, Lohse2016,Nakajima2016,zhang2017entangled, petrides:2018,Lee2018}. It has also been proposed to directly engineer 4DQH atomic or photonic systems~\cite{Price2015, Ozawa2016,Price2016} by using ``synthetic dimensions", where sets of internal states are coupled to mimic the connectivity of extra spatial dimensions~\cite{Boada2012,Celi:2012PRL, boada2015quantum, Mancini:2015Science, Stuhl:2015Science, luo2015quantum, Livi:2016PRL, yuan2016photonic, Ozawa2016, yuan2016photonic, ozawa2017synthetic, Price:2017PRA,An:2017SciAdv}. 

Unlike the 2DQH effect, the 4DQH effect is associated with several distinct topological symmetry classes~\cite{Chiu:2016RMP}, as the 2CN does not necessarily vanish with TRS. Indeed, this effect has previously been studied both for Class AII systems~\cite{IQHE4D,QiZhang,li:2012, Li:2013, li2013topological,Kraus2013}, where TRS for spinful fermions is preserved (as in the 2D quantum spin Hall effect), as well as for Class A systems~\cite{Kraus2013, Price2015, Ozawa2016, Price2016, zhang2017entangled, Lohse2018, Zilberberg2018}, where TRS is broken (as in the 2DQH effect). Physically, the models studied could be realised, for example, with spinful particles in non-Abelian gauge fields and spinless particles in magnetic fields, respectively. 

 In this paper, we propose how to realise a 4DQH lattice through designing the lattice connectivity with real-valued hopping amplitudes. The model has topological energy bands with non-zero 2CNs and so exhibits a 4DQH effect, but without requiring complicated (artificial) gauge fields and/or time-reversal symmetry breaking. As it relies on connectivity, the model could be implemented as a 4D network embedded in lower dimensions~\cite{jukic2013four}.  This model is also a minimal topological construction for Class AI, where TRS for {\it spinless} or {\it  bosonic} particles is preserved and for which nontrivial topological bands only arise in four dimensions and higher. In this class, the 2CN  only takes even integer values, meaning that surface states always come in pairs. Furthermore,  as all 1CNs vanish by TRS, the 2CN is always independent of lower-dimensional topological invariants, unlike in the (Class A) 4DQH systems recently probed experimentally, in which the 1CNs and the 2CN were related~\cite{Lohse2018,Zilberberg2018}. 
 
To introduce our proposal, we begin in Section~\ref{sec:minimal2D} by reviewing how to construct minimal 2D QH two-band models, such as the 2D Haldane model~\cite{Haldane:1988}. In Section~\ref{sec:minimal4D}, we then discuss the generalisation of these ideas into four spatial dimensions, in order to engineer Class AI energy bands with nontrivial 2CNs. We then propose a specific 4DQH lattice model in Section~\ref{sec:lattice4D}, which extends the widely-studied 2D honeycomb/brickwall lattice~\cite{goerbig2017dirac} into 4D. This provides a minimal 4D topological model through lattice connectivity. 

\section{Minimal 2DQH Two-Band Models} \label{sec:minimal2D}

In this section, we review simple ideas behind the construction of minimal 2D Quantum Hall two-band lattice models. This will provide us with a conceptual framework to later discuss minimal 4DQH four-band lattice models in Class AI. 

As topological invariants are integers, they do not change continuously but instead jump through topological phase transitions, where energy gaps between bands close and re-open. An intuitive approach for constructing topological models is therefore to focus on how such transitions can be induced by tuning the parameters of a Hamiltonian. Minimal 2DQH models can be constructed from two energy bands, as described generically by the momentum-space Hamiltonian~\cite{bernevig2013topological}:
\begin{eqnarray}
H ({\bf k}) = \varepsilon({\bf k}) \hat{I} + {\bf d} ({\bf k}) \cdot \bf{\sigma}, \label{eq:generic}
\end{eqnarray}
where $\hat{I}$ is the $2\times2$ identity, ${\bf k}$ is the momentum and $\bf{\sigma}$ is the vector of Pauli matrices. We focus on spinless models, where $\sigma$ represents a pseudo-spin degree of freedom, such as a sublattice basis. The two bands are given by: $E_\pm = \varepsilon({\bf k}) \pm \sqrt{{\bf d} ({\bf k}) \cdot {\bf d} ({\bf k})}$, where ${\bf d} ({\bf k})$ is a three-component vector and $\varepsilon({\bf k})$ is an overall energy shift, neglected without loss of generality hereafter.  
When the two bands are gapped, the topological 1CN (of the lower band) is given by~\cite{bernevig2013topological}: 
\begin{eqnarray}
\nu^{(1)}_- &=& \frac{1}{2 \pi} \int_{\text{BZ}} \Omega_- =  \frac{1}{4\pi} \int_{\text{BZ}} {d}^2 {\bf k} \epsilon^{abc} \hat{d}_a \partial_{k_x} \hat{d}_b  \partial_{k_y} \hat{d}_c  , \label{eq:chern}
\end{eqnarray}
where the integrals run over the 2D Brillouin zone (BZ), $ \epsilon^{abc} $ is the 3D Levi-Civita symbol, $ \hat{{\bf d}}={\bf d} / |{\bf d}|$ and $\Omega_-$ is the Berry curvature two-form of the lower band~\cite{Nakahara,Price2016}. The RHS is reached by finding $\Omega_-$ for the eigenstates of Eq.~\ref{eq:generic}. 
In this form, the 1CN of a two-band model is interpreted as the ``winding number"~\cite{bernevig2013topological}, counting how often $\hat{{\bf d}} ({\bf k})$ covers the unit Bloch sphere, $S^2$, in the BZ. 

As introduced above, the 1CN only changes via topological phase transitions, where the band-gap closes and re-opens. In the simplest case, the two bands touch, at the transition, at a set of isolated Dirac points in the BZ. Around each Dirac point, Eq.~\ref{eq:generic} can be expanded linearly such that, locally (up to a rotation), ${{\bf d}} ({\bf q}) \approx (v_x q_x, v_y q_y, M)$, where ${\bf q}$ is the momentum relative to the Dirac point, and $v_x(v_y)$ is the dispersion slope with respect to $q_x (q_y)$. The mass, $M$, smoothly tunes across the transition, as the Dirac point closes and re-opens as $M$ changes sign. 

Crucially, flipping the sign of $M$ also flips the sign of the Berry curvature (as $d_3\!\approx\!M$ in Eq.~\ref{eq:chern}). Indeed, it can be shown~\cite{bernevig2013topological} that each isolated Dirac point that closes and opens changes the 1CN by $\pm 1$. However, the sign of this change depends on the signs of the other two components, $d_1\!\approx\!v_x q_x$ and $d_2\!\approx\!v_y q_y$. If they have the same (opposite) sign, the Dirac point increases (decreases) the 1CN as $M$ goes from negative to positive. Whether a transition is topological then depends on how many Dirac points of each type there are. 

This argument has important consequences for the construction of simple 2D QH models. For spinless systems, TRS implies that there are equal numbers of Dirac points of both types in the BZ. This is because the spinless TRS operator is $\mathcal{T}\!=\!\mathcal{K}$, where $\mathcal{K}$ is complex conjugation, and so when TRS is present, $\mathcal{T} H ({\bf k}) \mathcal{T}^{-1} \!=\!  H (-{\bf k})$, then $d_1({\bf k})\!=\! d_1(-{\bf k})$, $d_2({\bf k})\!=\! -d_2(-{\bf k})$, $d_3({\bf k})\!=\! d_3(-{\bf k})$. A Dirac point at momentum ${\bf K}$ is therefore paired with another Dirac point of the opposite type (as $d_2$ must flip sign) at momentum $-{\bf K}$. These constraints also rule out unpaired Dirac points at TRS-invariant momenta, ${\bf k} \!=\! - {\bf k}$. Any transition that preserves TRS is then topologically trivial in 2D. Note that if $\sigma$ represents a real spin, ``fermionic" TRS requires that $\mathcal{T}^2=-1$ instead of $\mathcal{T}^2=+1$ as here; hence, the TRS operator is modified, and the two bands are only gapped if TRS is broken~\cite{bernevig2013topological}.  

To design a spinless topological model, the Dirac points in each pair need to be controlled separately by breaking TRS. This is beautifully illustrated by the Haldane model~\cite{Haldane:1988}, based on a 2D honeycomb lattice, such as graphene, or equivalently, a brickwall lattice, as in cold atom experiments~\cite{jotzu2014experimental}. Both lattices have two sites per unit cell, and can be modelled by a two-band Hamiltonian like Eq.~\ref{eq:generic}, where $\sigma$ is a sublattice basis~\cite{goerbig2017dirac}. If only nearest-neighbour hoppings are present, there is one pair of Dirac points in the BZ, which can be gapped out together by a momentum-independent mass, $M\sigma_3$. Physically, this corresponds to adding an energy offset between the two sites, breaking inversion symmetry and preserving TRS. 

In the Haldane model, TRS is broken by including complex next-nearest-neighbour hoppings~\cite{Haldane:1988}. These are designed such that, close to the two Dirac points, the local vector $d_3\!\approx\! M\pm M_1$, where $M_1$ depends on the geometry, complex hopping phase and amplitude. Then, one Dirac point closes and re-opens at $M\!=\!M_1$ and the other at $M\!=\!-M_1$, such that the model has a 1CN of $\pm1$ for $\! M \!<\! |M_1|$, as experimentally probed in cold atoms~\cite{jotzu2014experimental, flaschner2018observation}.    

\section{4D Class AI Topological Models} \label{sec:minimal4D}

We now show how extending the above ideas can lead to 4DQH models with nontrivial topological 2CNs, which do not require TRS-breaking or complicated gauge fields. In 4D, minimal QH systems can be constructed from four-band models of the form~\cite{QiZhang}: 
\begin{eqnarray}
H({\bf k}) = \varepsilon({\bf k}) \Gamma_0 + {\bf d}({\bf k}) \cdot {\bf \Gamma}, \label{eq:ham}
 \end{eqnarray}
where $\Gamma_0$ is the $4\times 4$ identity; ${\bf d}({\bf k})$ is a five-component vector; $\varepsilon({\bf k})$ is an overall energy shift, neglected without loss of generality hereafter; and $ {\bf \Gamma}$ is a vector of $4\times 4$ Dirac matrices, which are chosen as
\begin{eqnarray}
\Gamma_1 &= \left( \begin{array}{cccc} 0&0&1&0\\ 0&0&0&-1\\1&0&0&0\\0&-1&0&0\end{array} \right),\quad
&\Gamma_2 = \left( \begin{array}{cccc} 0&0&-i&0\\ 0&0&0&-i\\i&0&0&0\\0&i&0&0\end{array} \right), \nonumber \\
\Gamma_3 &=  \left( \begin{array}{cccc} 0&0&0&1\\ 0&0&1&0\\0&1&0&0\\1&0&0&0\end{array} \right),  \quad
&\Gamma_4 = \left( \begin{array}{cccc} 0&0&0&-i\\ 0&0&i&0\\0&-i&0&0\\i&0&0&0\end{array} \right),   \nonumber \\
\Gamma_5&=\left( \begin{array}{cccc} 1&0&0&0\\ 0&1&0&0\\0&0&-1&0\\0&0&0&-1\end{array} \right) &.  \qquad \label{eq:diracmatrices}
\end{eqnarray}
Note that, unlike a two-band model, the decomposition in Eq.~\ref{eq:ham} is not generic, and the energy bands are doubly-degenerate: $E_\pm = \varepsilon({\bf k}) \pm \sqrt{{\bf d} ({\bf k}) \cdot {\bf d} ({\bf k})}$. The 2CN for the lower pair of bands is~\cite{Nakahara,QiZhang}: 
\begin{eqnarray}
\nu^{(2)}_- &=&  \frac{1}{8 \pi^2} \int_{\text{BZ}} \text{tr}(\Omega_-\wedge \Omega_-) ,\nonumber \\
&=&  \frac{3}{8\pi^2} \int_{\text{BZ}} {d}^4 {\bf k} \epsilon^{abcde} \hat{d}_a \partial_{k_x} \hat{d}_b  \partial_{k_y} \hat{d}_c
 \partial_{k_z} \hat{d}_d
  \partial_{k_w} \hat{d}_e , \label{eq:secondchern}
\end{eqnarray}
where the trace runs over the Berry curvature wedge product of the lower band pair. Here, the integral is over the 4D BZ, with $ \epsilon^{abcde} $ being the 5D Levi-Civita symbol, and $ \hat{{\bf d}}={\bf d} / |{\bf d}|$ as above. In such a four-band model, the 2CN is again a ``winding number", but now counting how often $\hat{{\bf d}} ({\bf k})$ covers the unit sphere, $S^4$, across the 4D BZ.

As in 2D, the topological invariant of a Bloch band changes via topological phase transitions where band gaps close and re-open. In the simplest case, the four bands touch at an isolated set of Dirac points in the BZ, around each of which ${{\bf d}} ({\bf q}) \!\approx\! (v_x q_x, v_y q_y, v_z q_z, v_w q_w, M)$, where $v_{z} (v_{w})$ is the dispersion slope with respect to $q_z (q_w)$. As before, the mass, $M$, smoothly tunes across the transition, with the integrand of Eq.~\ref{eq:secondchern} flipping sign as $d_5\!\approx\!M$ changes sign. Each isolated point that closes and opens changes the 2CN by $\pm1$~\cite{QiZhang}. Again, this divides the Dirac points into two types; the first (second) type has an even (odd) number of minus signs within the other components $\{d_1, d_2, d_3, d_4\}$ such that the 2CN increases (decreases) as $M$ goes from negative to positive values.

Importantly, in 4D, preserving TRS for spinless systems {\it does not imply} equal numbers of the two types of Dirac points. This is because when spinless or bosonic TRS is present, $d_1({\bf k})\!=\! d_1(-{\bf k})$, $d_3({\bf k})\!=\! d_3(-{\bf k})$ and  $d_5({\bf k})\!=\! d_5(-{\bf k})$ are even, while $d_2({\bf k})\!=\! -d_2(-{\bf k})$ and $d_4({\bf k})\!=\! -d_4(-{\bf k})$ are odd. Then, a Dirac point at ${\bf K}$ is again paired with another Dirac point at $-{\bf K}$, but now these Dirac points are of the same type, as both $d_2$ and $d_4$ flip sign. As a result, each TRS pair of Dirac points changes the 2CN by $\pm2$ across a transition. For our  construction, the 2CN can only be an even integer, and in fact, this is a general property~\cite{Chiu:2016RMP} of Class AI in 4D. 

As well as supporting a bulk 4DQH response, a non-trivial 2CN implies the existence of 3D topological surface states, as there is a one-to-one correspondence between the bulk topological invariant and the number of topological surface states~\cite{RMP_TI, RMP_TI2,bernevig2013topological,QiZhang}. For Class AI models, the fact that the 2CN takes even integer values means that these surface states always come in pairs. In the 3D BZ of our model, these correspond to pairs of Weyl points with the same chirality. Moreover, as the 1CNs always vanish due to TRS, both the bulk 4DQH response and the 3D topological surface states are intrinsically related to the 4D topological 2CN, which is independent of all lower-dimensional topological invariants. This is in contrast to the (Class A) 4DQH models recently probed experimentally~\cite{Lohse2018, Zilberberg2018}, in which the 2CN was not fully independent of the lower-dimensional 1CNs.  

If instead we had considered particles with half-integer spin, then we would have had to impose fermionic TRS, which, as discussed above, modifies the TRS operator and constraints. This leads to the well-known construction of a four-band (Class AII) 4DQH model~\cite{QiZhang}, which also has Weyl-point surface states. For fermionic TRS, unpaired Dirac points are allowed at TRS-invariant momenta, meaning that the 2CN can change by $\pm 1$ in general, and so the 2CN can be any integer. As for the spinless system discussed above, the presence of TRS means that 1CNs vanish, and so the 2CN is independent of lower-dimensional topological properties. However, it may be challenging to realise the Class AII 4DQH model experimentally, as it would describe, for example, a lattice of spinful particles in spatially-varying, spin-dependent gauge fields. Instead, as we discuss below, a suitable (Class AI) four-band model can be engineered for spinless particles by simply exploiting lattice connectivity, with purely real hopping amplitudes.

Finally, we also note that while for 4DQH band models the integral in Eq.~\ref{eq:secondchern} runs over the crystal momenta in the 4D BZ, the 2CN can itself be defined over any suitable closed 4D manifold. In this context, a nontrivial 2CN was recently simulated in the parameter space of an atomic Bose-Einstein condensate, where four hyperfine atomic states were cyclically coupled by external electromagnetic fields~\cite{Sugawa:2016arXiv}. In this case, time-reversal symmetry was also preserved, leading to the realization of two-fold degenerate states, characterised in parameter space by a non-zero 2CN and vanishing 1CNs.     

\section{Proposal for a 4D Topological Lattice through connectivity} \label{sec:lattice4D}

Inspired by the 2D Haldane model, our 4D proposal extends the honeycomb/brickwall lattice into 4D. As introduced above, these lattices are topologically-equivalent, having two sites per unit cell and a single pair of Dirac points in the BZ. Hereafter, we choose to focus on the brickwall geometry, but note that similar arguments apply to the honeycomb geometry.   

\begin{figure*}[!]
\includegraphics[width=0.95\linewidth]{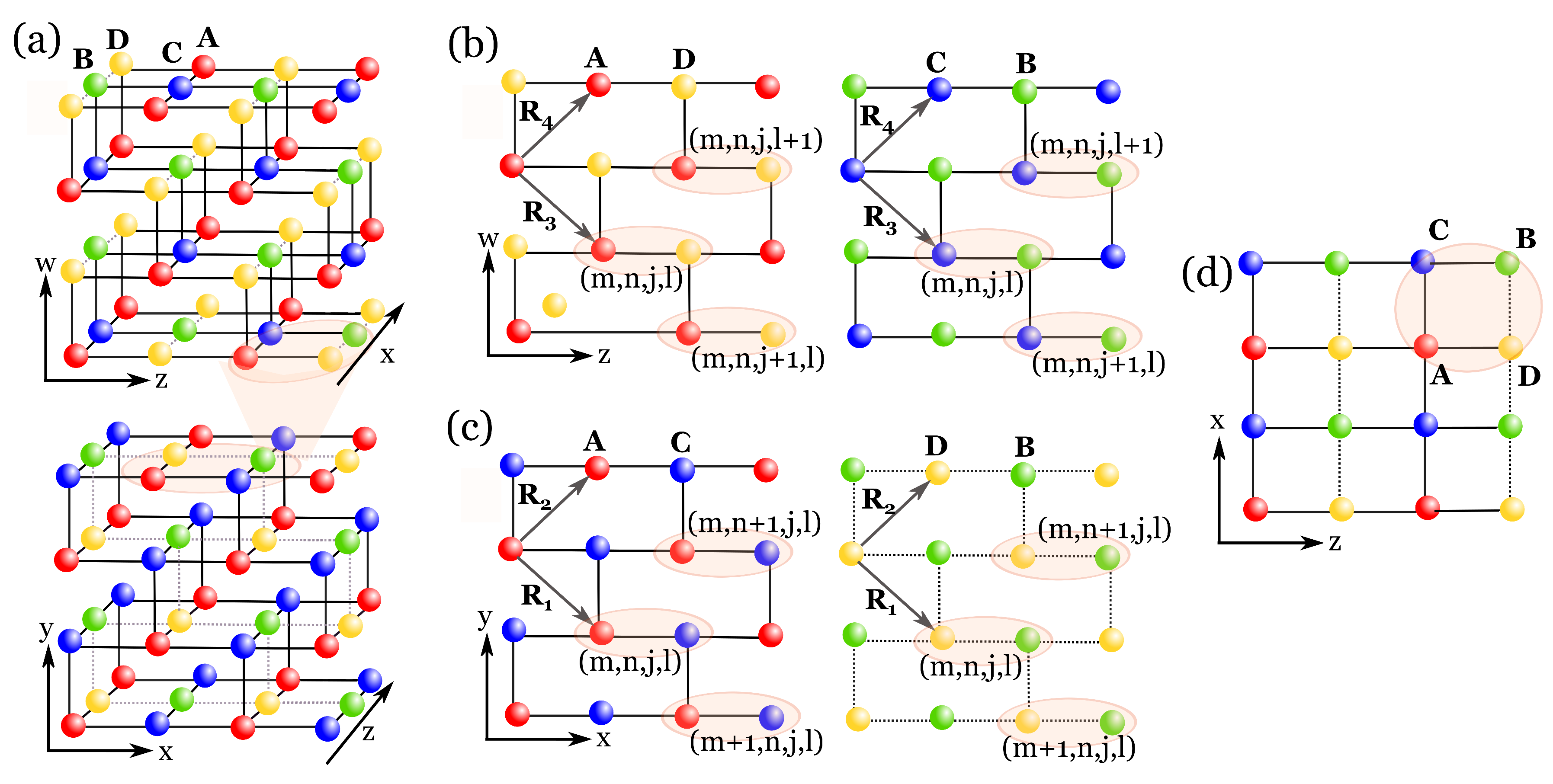} 
\caption{\label{fig1}  (a) Schematic of the 4D brickwall lattice with four sites per unit cell. Solid and dotted lines denote hoppings with real amplitudes $J$ and $-J$ respectively. Cuts of the 4D brickwall lattice in (b) the $z$-$w$ plane showing $A$-$D$ hoppings and $C$-$B$ hoppings, and (c) the $x$-$y$ plane showing $A$-$C$ hoppings and $B$-$D$ hoppings respectively. The set of lattice vectors are marked: ${\bf R}_1 = (1, -1,0,0)$, ${\bf R}_2 = (1, 1,0,0)$, ${\bf R}_3 = (0, 0,1,-1)$ ${\bf R}_4 =(0, 0,1,1)$, and the indices $(m,n,j,l)$ label a given unit cell with respect to these lattice vectors. (e) Cut of the 4D brickwall lattice in the $x$-$z$ plane. In all panels, the unit cell structure is highlighted by pale pink shading. }	
\end{figure*}

To realise a four-band model like Eq.~\ref{eq:ham}, we start by constructing a 4D lattice [see Fig.~\ref{fig1}(a)], with a four-site unit cell and the connectivity of a 2D brickwall lattice in both $x\!-\!y$ and $z\!-\!w$ planes. As shown here, our lattice has four sites, denoted by $(A, B, C, D)$. The corresponding set of lattice vectors are: 
${\bf R}_1 = (1, -1,0,0)$, ${\bf R}_2 = (1, 1,0,0)$, ${\bf R}_3 = (0, 0,1,-1)$ ${\bf R}_4 =(0, 0,1,1)$, with $a=1$ being the distance between any two nearest-neighbour lattice sites.  The full real-space tight-binding Hamiltonian is given by
\begin{eqnarray}
H&=& H_x + H_y + H_z + H_w + H_{{0}} \nonumber ,\end{eqnarray}
with hopping terms along each direction as:
\begin{eqnarray} 
H_x &=& J  \sum_{m,n,j,l}  (c^\dagger_{m,n,j,l} a_{m,n,j,l} + a^\dagger_{m+1,n+1,j,l} c_{m,n,j,l} \nonumber \\ &&- b^\dagger_{m,n,j,l} d_{m,n,j,l} - d^\dagger_{m+1,n+1,j,l} b_{m,n,j,l} + \rm{h.c}) \nonumber \\
H_y &=& J \sum_{m,n,j,l}  (c^\dagger_{m-1,n,j,l} a_{m,n,j,l} - b^\dagger_{m-1,n,j,l} d_{m,n,j,l}  + \rm{h.c} ) \nonumber \\
H_z &=& J \sum_{m,n,j,l}  (d^\dagger_{m,n,j,l} a_{m,n,j,l} + a^\dagger_{m,n,j+1,l+1} d_{m,n,j,l}  \nonumber \\ &&+ b^\dagger_{m,n,j,l} c_{m,n,j,l} + c^\dagger_{m,n,j+1,l+1} b_{m,n,j,l} + \rm{h.c} ) \nonumber \\
H_w&=& J \sum_{m,n,j,l} (d^\dagger_{m,n,j-1,l} a_{m,n,j,l}+b^\dagger_{m,n,j-1,l} c_{m,n,j,l}  + \rm{h.c} )  \nonumber
\end{eqnarray}
and with on-site terms: 
\begin{eqnarray}
H_{{0}} &=& M \sum_{m,n,j,l} (a^\dagger_{m,n,j,l} a_{m,n,j,l}  + b^\dagger_{m,n,j,l} b_{m,n,j,l}  \nonumber \\ && -c^\dagger_{m,n,j,l} c_{m,n,j,l} -d^\dagger_{m,n,j,l} d_{m,n,j,l} ). \nonumber
\end{eqnarray}
Here, $J$ is the hopping amplitude, which we can take to be real-valued, and $M$ is an energy offset between the $A, B$ and $C, D$ sites. The index $(m,n,j,l)$ indicates a particular unit cell with respect to lattice vectors $({\bf R}_1,{\bf R}_2, {\bf R}_3, {\bf R}_4)$, and we have introduced the operators $\alpha_{m,n,j,l}$ ($\alpha^\dagger_{m,n,j,l}$) which annihilate (create) a particle on an $\alpha$-site in the $(m,n,j,l)$ unit cell. By Fourier-transforming these operators as: 
 \begin{eqnarray}
 a_{m,n,j,l} &\propto&  \sum_{{\bf k}} a_{\bf k} e^{- i {\bf k}\cdot [m {\bf R}_1+ n {\bf R}_2+ j {\bf R}_3+ l {\bf R}_4]} \nonumber \\ 
  b_{m,n,j,l} &\propto&  \sum_{{\bf k}} b_{\bf k} e^{- i {\bf k}\cdot [(m+\frac{1}{2}) {\bf R}_1+ (n+\frac{1}{2}) {\bf R}_2+ (j+\frac{1}{2}) {\bf R}_3+ (l+\frac{1}{2})  {\bf R}_4]} \nonumber \\ 
   c_{m,n,j,l} &\propto&   \sum_{{\bf k}} c_{\bf k} e^{- i {\bf k} \cdot  [(m+\frac{1}{2}) {\bf R}_1+ (n+\frac{1}{2}){\bf R}_2+ j  {\bf R}_3+ l  {\bf R}_4]}  \nonumber \\ 
   d_{m,n,j,l} &\propto&  \sum_{{\bf k}} d_{\bf k} e^{- i {\bf k} \cdot  [m {\bf R}_1+ n  {\bf R}_2+ (j+\frac{1}{2}) {\bf R}_3+(l+\frac{1}{2}) {\bf R}_4]}  \label{eq:fourier}
 \end{eqnarray}
where the sum runs over all momenta in the BZ, we find the momentum-space Hamiltonian (see Appendix A): 
\begin{eqnarray}
H({\bf k}) &= J& \left[ (2 \cos k_x + \cos k_y) \Gamma_1 +  \sin k_y \Gamma_2 \right. \nonumber \\ 
&& \left.+ (2 \cos k_z + \cos k_w) \Gamma_3 +  \sin k_w \Gamma_4\right] + M \Gamma_5 . \qquad \label{eq:model}
\end{eqnarray} 
Note that the real-space hoppings between $B$ and $D$ sites need to have an opposite sign compared to other hoppings, as indicated in Fig.~\ref{fig1}(a), in order to realise the required $\Gamma$ matrix structure. This model is a specific example of the general form given in Eq.~\ref{eq:ham}, corresponding to a four-band model. 

\begin{figure}[t!]
$\begin{array}{cc}
	(a) \includegraphics[width=0.43\linewidth]{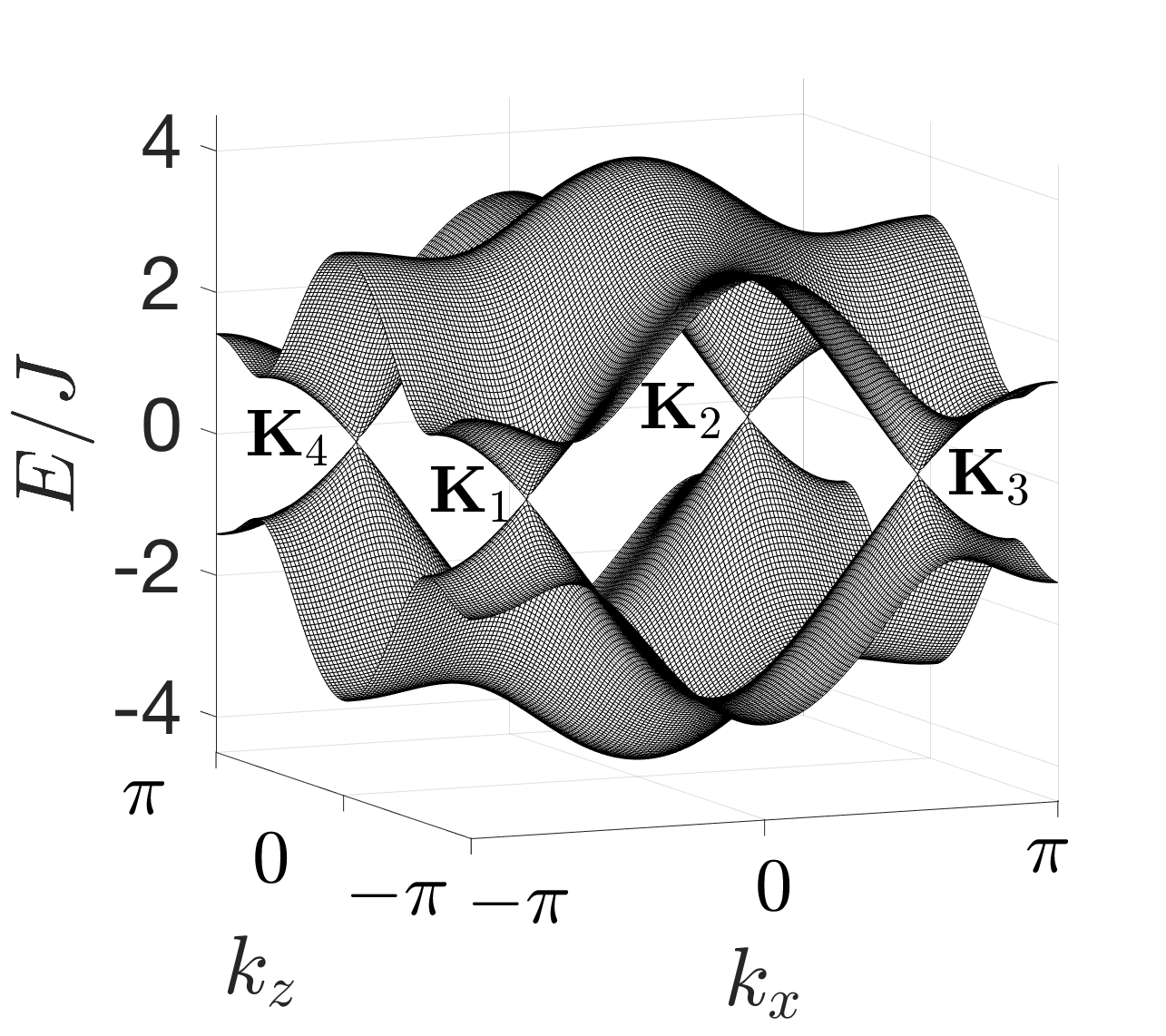} 
&
(b) \includegraphics[width=0.45\linewidth]{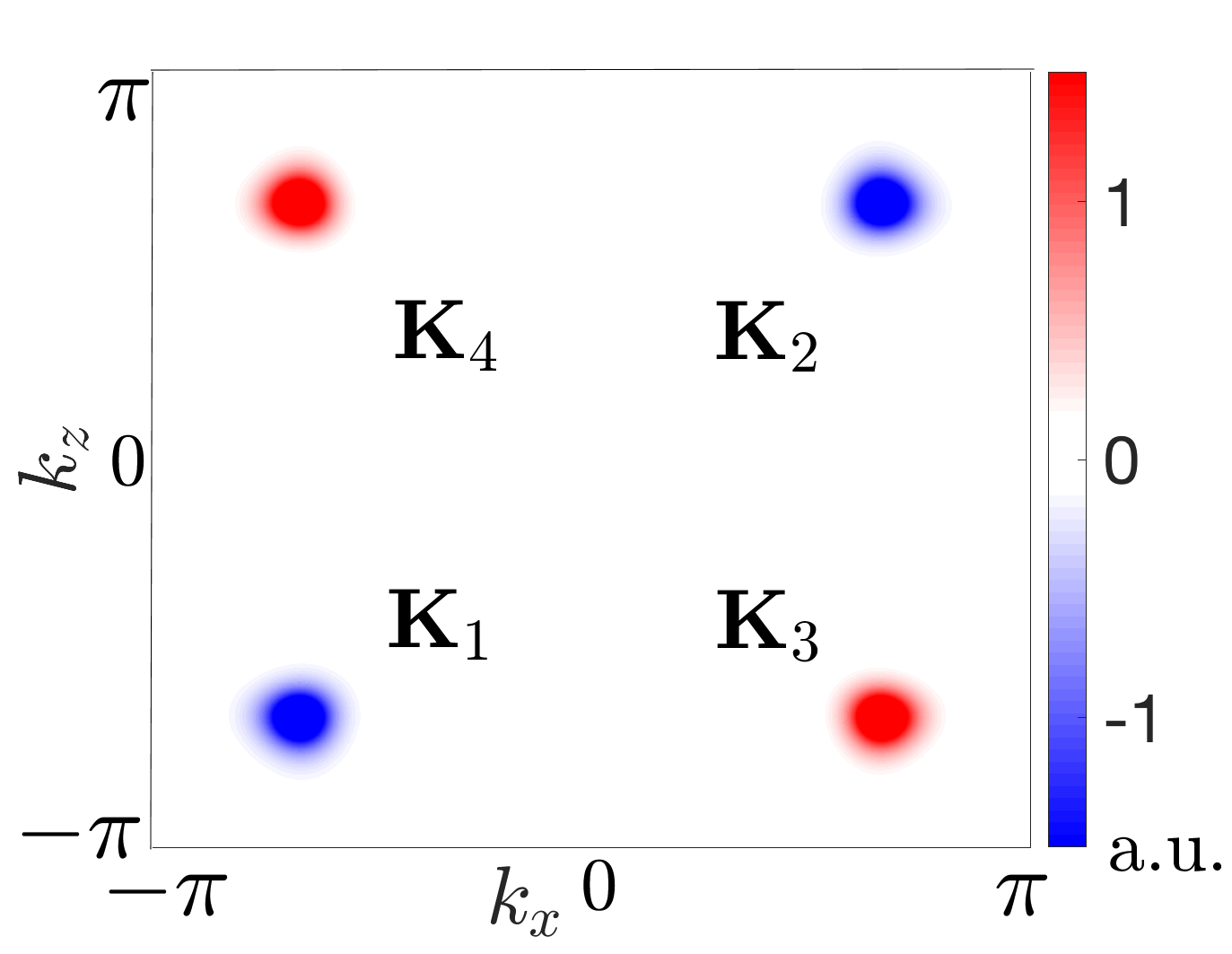} 
	\end{array}$
	\caption{\label{fig2} (a) The doubly-degenerate energy bands (Eq.~\ref{eq:model}) for $k_y=k_w=0$ and $M=0$, with the 4D Dirac points labelled. (b) When $M\neq 0$, the Dirac points are gapped and the integrand of Eq.~\ref{eq:secondchern} is nontrivial, as shown here for $k_y\!=\!k_w\!=\!0$ and $M\!=\!-J/2$. However, this lattice is topologically trivial as the two pairs of Dirac points contribute in opposite senses to the 2CN. }	
\end{figure}

When $M=0$, this model has four 4D Dirac points in the BZ, as shown in Fig.~\ref{fig2}(a). The points at ${\bf K}_{1,2}=(\mp 2\pi/3, 0, \mp 2\pi/3, 0)$ are a time-reversal pair of the first type, while those at ${\bf K}_{3,4}=(\pm 2\pi/3, 0, \mp 2\pi/3, 0)$ are a pair of the second type. Therefore, this model is still topologically trivial, as shown, for example, in Fig.~\ref{fig2}(b), for a cut at $k_y\!=\!k_w\!=\!0$ and $M\!=\!-J/2$, where the contributions to the 2CN (Eq.~\ref{eq:secondchern}) clearly cancel out for the two pairs.

\begin{figure}[!]
$\begin{array}{cc}

			(a) \includegraphics[width=0.4\linewidth]{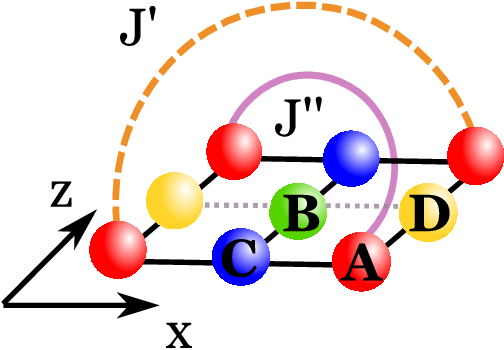} & (b) \includegraphics[width=0.43\linewidth]{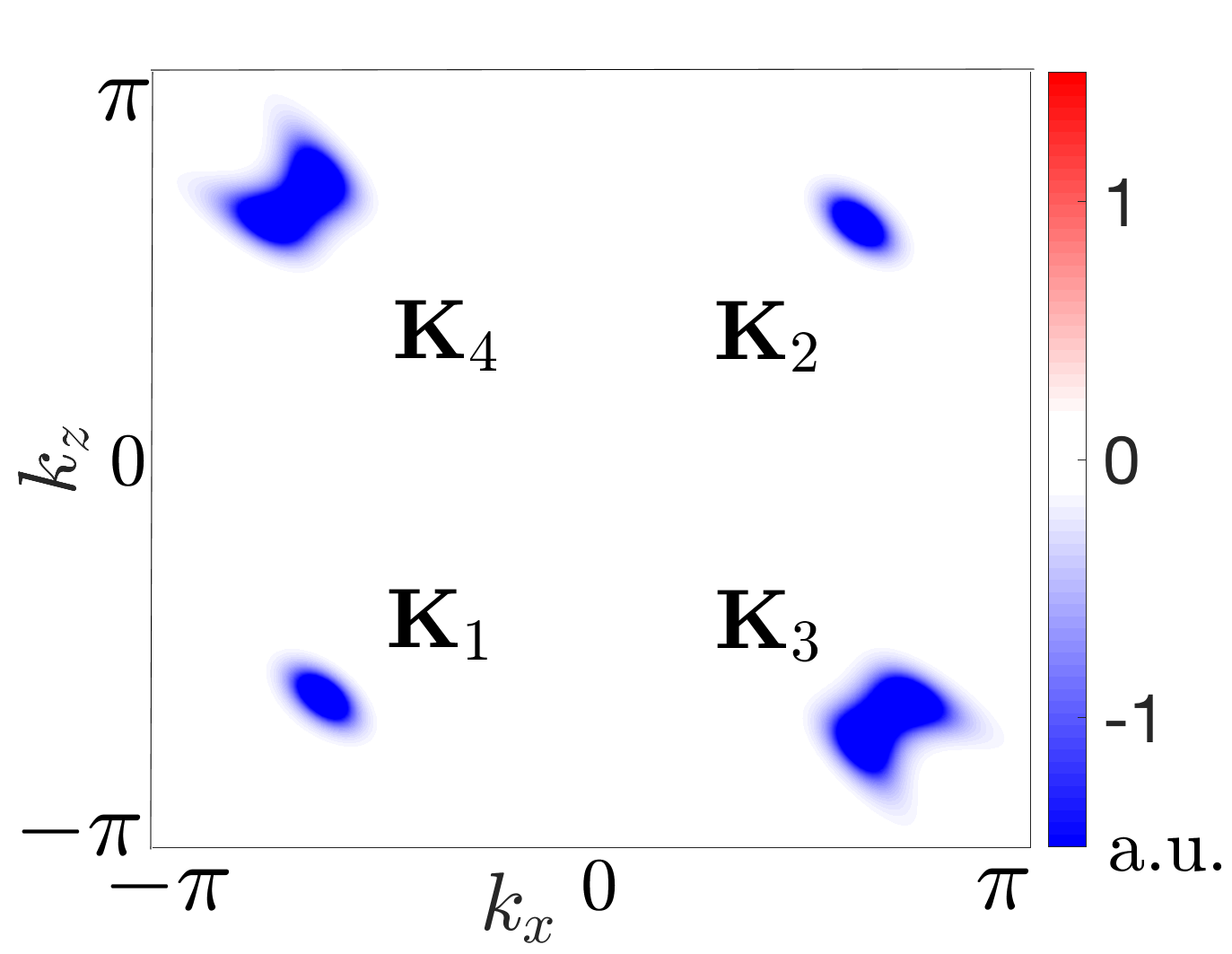} \\

	\\	(c) \includegraphics[width=0.43\linewidth]{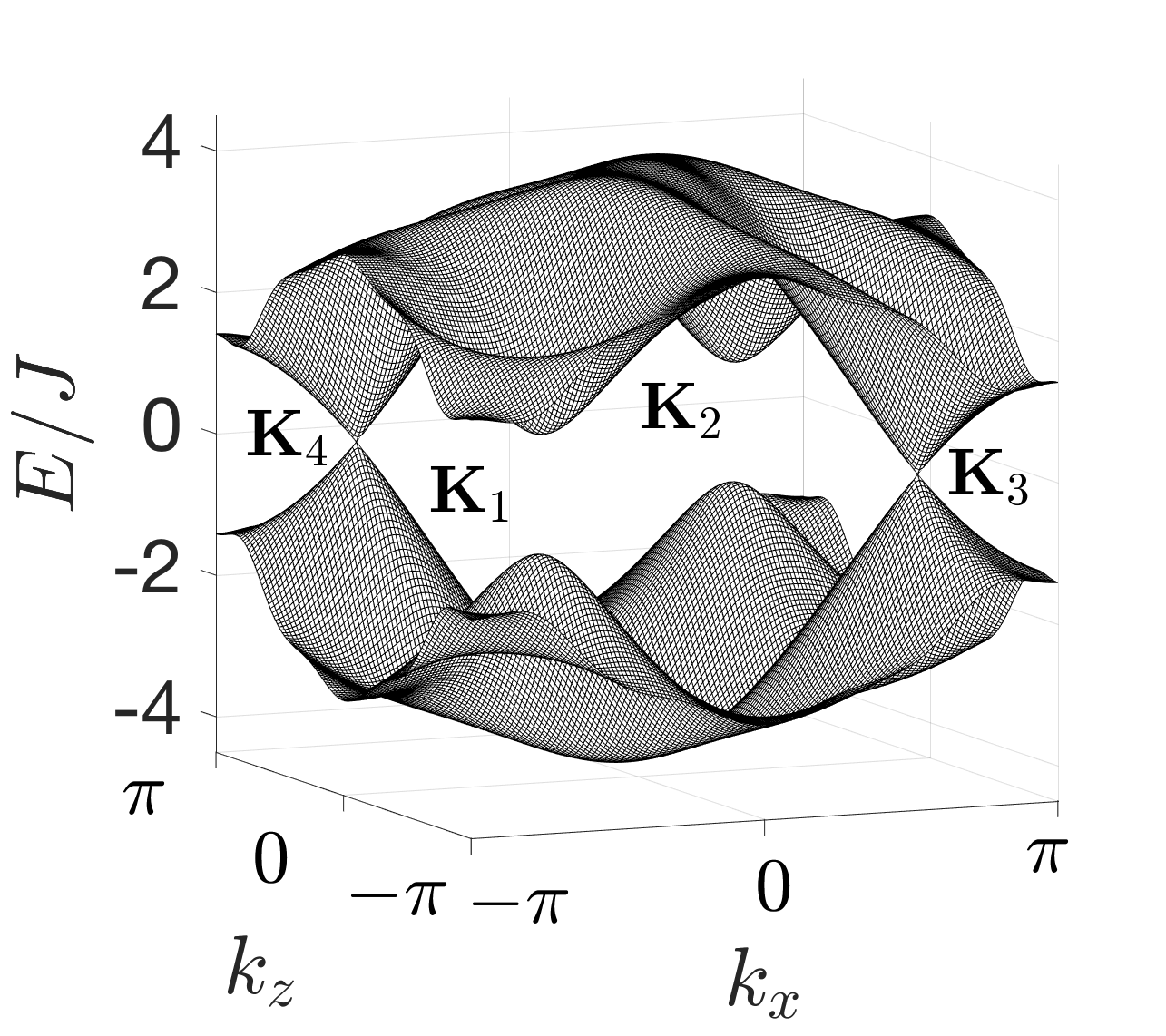} &(d) \includegraphics[width=0.43\linewidth]{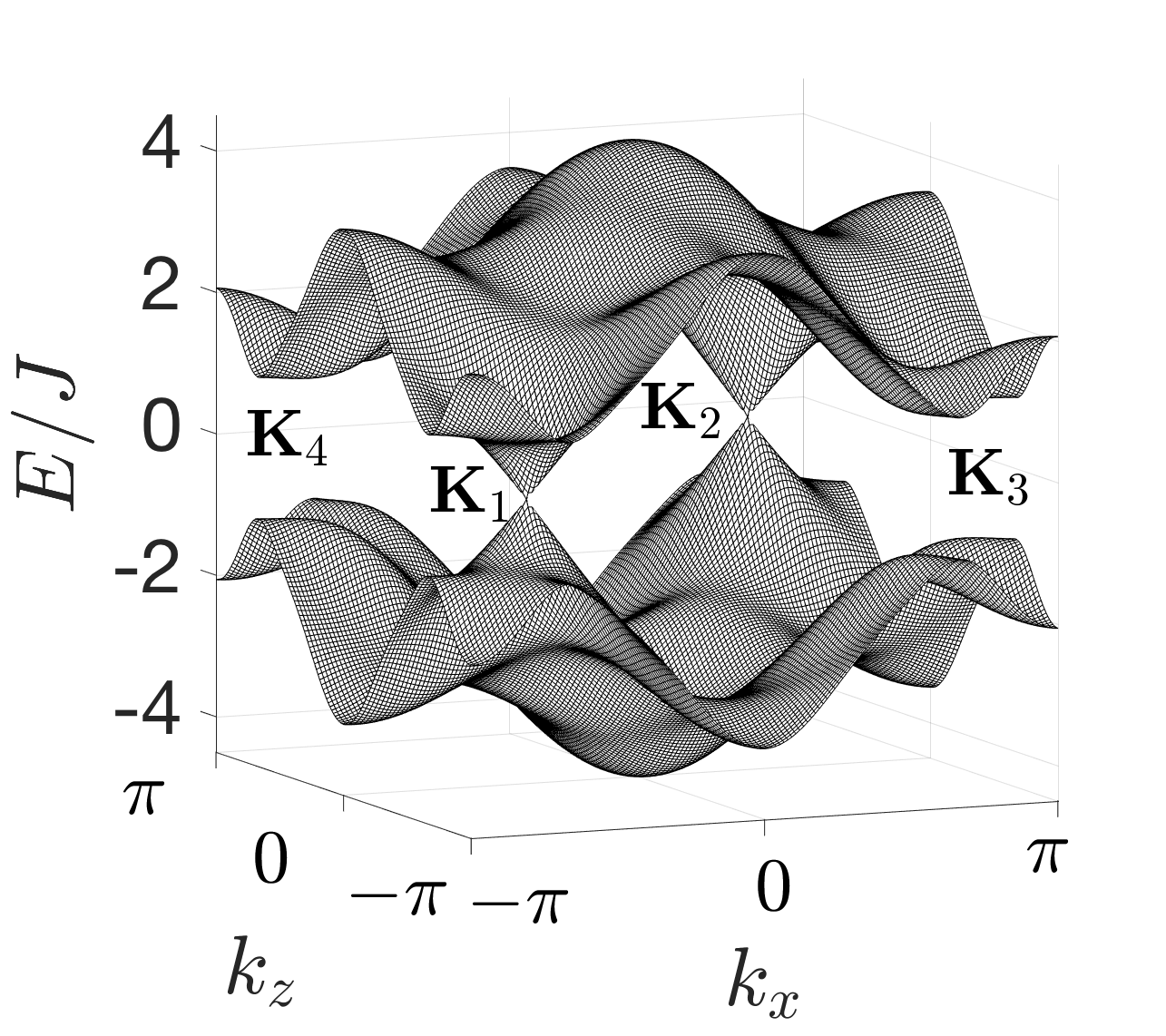} 
	\end{array}$
	\caption{\label{fig3} (a) Example of longer-range hoppings (Eq.~\ref{eq:mass}) that can make the lattice [Fig.~\ref{fig1}(a)] topologically nontrivial. (b) The integrand of Eq.~\ref{eq:secondchern} with $k_y\!=\!k_w\!=\!0$, $m\!=\!- J/2$, $J'\!=\!J/2$ and $J''\!=\!0$, showing that the two pairs now contribute to the 2CN in the same sense, giving a total 2CN of $-2$. (c)\&(d) The energy dispersion at the topological phase transitions, corresponding, for the parameters above, to (c) $\!m\!=\!-J$ and (d) $m\!=\!J/2$, showing that there is only one pair of Dirac points at each transition. 
	}	
\end{figure}

As in the 2D Haldane model, another ingredient is needed to separate out the two types of Dirac points and engineer topological bands. In particular, we need a mass-like term, proportional to $\Gamma_5$, which distinguishes between the two pairs of Dirac points. As an example, we consider long-range hoppings in the $x-z$ plane along ${\bf r}' = (\pm 2 a, 0, \pm 2a, 0)$ and ${\bf r}''= (\pm 2 a, 0, \mp 2a, 0)$ (e.g. see Fig.~\ref{fig3}(a)). In terms of the tight-binding real-space model, this would correspond to adding terms: 
\begin{widetext}
\begin{eqnarray}
H_{\text{l}}& =&     J' \sum_{m,n,j,l} ( a^\dagger_{m+1,n+1,j+1,l+1} a_{m,n,j,l} +b^\dagger_{m+1,n+1,j+1,l+1} b_{m,n,j,l} -c^\dagger_{m+1,n+1,j+1,l+1} c_{m,n,j,l} -d^\dagger_{m+1,n+1,j+1,l+1} d_{m,n,j,l} + \rm{h.c} ) \nonumber \\
&&+ J'' \sum_{m,n,j,l} ( a^\dagger_{m+1,n+1,j-1,l-1} a_{m,n,j,l} +b^\dagger_{m+1,n+1,j-1,l-1} b_{m,n,j,l} -c^\dagger_{m+1,n+1,j-1,l-1} c_{m,n,j,l} -d^\dagger_{m+1,n+1,j-1,l-1} d_{m,n,j,l} + \rm{h.c} ) \nonumber
\end{eqnarray}
\end{widetext}
where we have allowed for the hoppings along ${\bf r}' = (\pm 2 a, 0, \pm 2a, 0)$ to have amplitude $J'$ and those along  ${\bf r}'' = (\pm 2 a, 0, \mp 2a, 0)$ to have amplitude $J''$.  Applying Eq,~\ref{eq:fourier} as above, the long-range hoppings lead to a momentum-space Hamiltonian of the form  (see Appendix A):
\begin{eqnarray}
H'({\bf k}) = \left[ 2 J' \cos (2 k_x\!+ 2 k_z\! ) + 2 J'' \cos (2 k_x\!- 2 k_z\! )\right]  \Gamma_5, \qquad  \label{eq:mass}
\end{eqnarray}
Note that both $J'$ and $J''$ can be taken to be real-valued, however, the hoppings from $A\!\rightarrow A$, $B\!\rightarrow B$ should have different signs to those from $C\!\rightarrow C$, $D\!\rightarrow D$, in order to get the required matrix structure in this momentum-space equation. 

As a result of these additional terms, the first pair of 4D Dirac cones closes at $M\!=\!J'\!-\!2J''$ and the second at $M\!=\!J''\!-\!2J'$ [see Fig.~\ref{fig3}(c)\&(d)]. Provided that $J' \!\neq \!J''$, these are topological transitions; for example, if $J''\!=\!0$ and $J'\!>\!0$, this model has a 2CN of -2 for $-2J' \!< \!M\! <\! J'$, and is trivial otherwise, as can also be confirmed numerically~\cite{Mochol2018}. Note that the above terms preserve TRS and so all 1CNs vanish by symmetry. Adding TRS-breaking terms will separate the Dirac points within a pair; this can give a 4D QH model, but in Class A where the 1CNs can also be non-zero~\cite{Kraus2013, Price2015, Lohse2018, Zilberberg2018}. 

We emphasise that the above is only one choice of long-range hoppings that will lead to topological bands. Indeed, it is clear that all that is required are hoppings between alike sites chosen such that the effective mass-term in momentum-space is proportional to $\Gamma_5$ and has a momentum-dependence such that it distinguishes between the first Dirac pair at ${\bf K}_1$ and ${\bf K}_2$ as compared to the second pair at ${\bf K}_3$ and ${\bf K}_4$. Other examples of appropriate terms could include: (1) hoppings along ${\bf r}''' = (a, a, 2a, 0)$ and similar, leading to momentum-space terms $\propto \cos (k_x+k_y +2 k_z)\Gamma_5$ etc, or (2) hoppings along ${\bf r}'''' = (a, a, a, a)$ and similar, leading to momentum-space terms $\propto \cos (k_x+k_y + k_z+k_w)\Gamma_5$ etc. In each case, a suitable design of these hoppings will lead to a similar topological phase diagram that has a topological phase with a 2CN of $|2|$ within certain parameters, and a trivial topological phase otherwise. As there is considerable freedom therefore in choosing the long-range hopping terms, the most suitable choice may depend on the specific experimental implementation. 

In practice, there may also be other long-range hopping terms present experimentally which are not of the desired type. However, the topological phase of this model will be robust, provided that these unwanted terms are sufficiently small. We note that those terms which cannot be expressed in terms of the five $\Gamma$ matrices introduced above can also break the double-degeneracy of the energy bands. While this may complicate the simple picture for counting Dirac points, the 2CN can still be calculated numerically according to the algorithm of Ref.~\onlinecite{Mochol2018}.

\section{Conclusions} \label{sec:conc}

In this paper, we have reviewed the construction of minimal 2DQH models, and extended these ideas to propose 4DQH systems with spinless TRS. We have shown that such a 4D topological system could be engineered by controlling the lattice connectivity, while requiring only real-valued positive and negative hoppings. This provides a new way to realise the 4DQH effect which does not rely on either time-reversal symmetry breaking and/or complicated gauge fields. This also provides a minimal topological model for Class AI, which describes spinless or bosonic models with TRS and which is topologically-trivial in lower dimensions. This work opens the way towards the experimental exploration of a higher-dimensional topological systems by controlling the lattice connectivity. \\

{\textbf{Note Added}:} In preparation of this manuscript, we became aware of a recent proposal for an eight-band 4D crystalline topological insulator, which has bosonic TRS~\cite{shao2018four}, but which is instead topologically-protected by reflection symmetry and which relies on spin-orbit couplings. Since this proposal was put on arXiv, it has been experimentally implemented in electric circuits~\cite{wang2020circuit}. Theoretical proposals have also been made for electric circuits to realise a different spinless (Class AI) 4DQH model~\cite{yu2019genuine}, to simulate $n$th-Chern-number insulators~\cite{ezawa2019electric} and to image nodal boundary Seifert surfaces in 4D circuits~\cite{li2019emergence}.\\

\begin{acknowledgments}

I am grateful to Tomoki Ozawa for inspiration and many helpful comments. I also thank Iacopo Carusotto, Ben McCanna and Cristiane Morais Smith for interesting and useful discussions. This work was supported by funding from the Royal Society via grants UF160112, RGF/EA/180121 and RGF/R1/180071. %

\end{acknowledgments}

\appendix
\section{Derivation of Momentum-Space Hamiltonians}~\label{sec:hopping}

In this Appendix, we provide additional detailed steps in the derivation of the momentum-space Hamiltonians presented in Section~\ref{sec:lattice4D}. Firstly, as stated in the main text, the real-space tight-binding Hamiltonian including up to nearest-neighbour hoppings is given by
\begin{eqnarray}
H&=& H_x + H_y + H_z + H_w + H_{{0}} \nonumber ,\end{eqnarray}
with hopping terms along each direction as:
\begin{eqnarray} 
H_x &=& J  \sum_{m,n,j,l}  (c^\dagger_{m,n,j,l} a_{m,n,j,l} + a^\dagger_{m+1,n+1,j,l} c_{m,n,j,l} \nonumber \\ &&- b^\dagger_{m,n,j,l} d_{m,n,j,l} - d^\dagger_{m+1,n+1,j,l} b_{m,n,j,l} + \rm{h.c}) \nonumber \\
H_y &=& J \sum_{m,n,j,l}  (c^\dagger_{m-1,n,j,l} a_{m,n,j,l} - b^\dagger_{m-1,n,j,l} d_{m,n,j,l}  + \rm{h.c} ) \nonumber \\
H_z &=& J \sum_{m,n,j,l}  (d^\dagger_{m,n,j,l} a_{m,n,j,l} + a^\dagger_{m,n,j+1,l+1} d_{m,n,j,l}  \nonumber \\ &&+ b^\dagger_{m,n,j,l} c_{m,n,j,l} + c^\dagger_{m,n,j+1,l+1} b_{m,n,j,l} + \rm{h.c} ) \nonumber \\
H_w&=& J \sum_{m,n,j,l} (d^\dagger_{m,n,j-1,l} a_{m,n,j,l}+b^\dagger_{m,n,j-1,l} c_{m,n,j,l}  + \rm{h.c} )  \nonumber
\end{eqnarray}
and with on-site terms: 
\begin{eqnarray}
H_{{0}} &=& M \sum_{m,n,j,l} (a^\dagger_{m,n,j,l} a_{m,n,j,l}  + b^\dagger_{m,n,j,l} b_{m,n,j,l}  \nonumber \\ && -c^\dagger_{m,n,j,l} c_{m,n,j,l} -d^\dagger_{m,n,j,l} d_{m,n,j,l} ). \nonumber
\end{eqnarray}
To proceed, we Fourier-transform each operator [using Eq.~\ref{eq:fourier}], such that the Hamiltonian can be written as
\begin{eqnarray} 
H = \sum_{\bf k}  
\left( \begin{array}{cccc} a^\dagger_{\bf k}&b^\dagger_{\bf k}&c^\dagger_{\bf k}&d^\dagger_{\bf k}\end{array} \right)
H ({\bf k})  \left( \begin{array}{c} a_{\bf k}\\b_{\bf k}\\c_{\bf k}\\d_{\bf k}\end{array} \right), 
\end{eqnarray}
where
\begin{eqnarray} 
H ({\bf k}) = H_x ( {\bf k}  ) + H_y ( {\bf k}  ) + H_z ( {\bf k}  ) + H_w ( {\bf k}  ) + H_0 ( {\bf k}  ). \nonumber 
\end{eqnarray} 
This procedure leads to the following expressions:
\begin{eqnarray}
H_x ( {\bf k}  ) &=&  
\left( \begin{array}{cccc} 0&0&2 J \cos k_x &0\\ 0&0&0&-2 J \cos k_x \\2 J \cos k_x &0&0&0\\0&-2 J \cos k_x &0&0\end{array} \right), \nonumber 
\end{eqnarray}
\begin{eqnarray}
H_ y  ( {\bf k}  )&= &  \left( \begin{array}{cccc} 0&0&J e^{-i k_y}&0\\ 0&0&0&-J e^{i k_y}\\J e^{i k_y}&0&0&0\\0&-J e^{-i k_y}&0&0\end{array} \right), \nonumber
\end{eqnarray}
\begin{eqnarray}
H_z  ( {\bf k}  )&=& 
\left( \begin{array}{cccc} 0&0&0&2 J \cos k_z \\ 0&0&2 J \cos k_z &0\\0&2 J \cos k_z&0&0\\2 J \cos k_z &0&0&0\end{array} \right) ,  \nonumber 
\end{eqnarray}
\begin{eqnarray}
H_w  ( {\bf k}  )&=& 
 \left( \begin{array}{cccc} 0&0&0&J e^{-i k_w}\\ 0&0&J e^{i k_w}&0\\0&J e^{-i k_w}&0&0\\J e^{i k_w}&0&0&0\end{array} \right),  \nonumber 
\end{eqnarray}
\begin{eqnarray}
H_{0}  ( {\bf k}  ) &= &
 \left( \begin{array}{cccc} M&0&0&0\\ 0&M&0&0\\0&0&-M&0\\0&0&0&-M\end{array} \right), \nonumber
\end{eqnarray}
Using the Dirac matrices defined in Eq.~\ref{eq:diracmatrices}, the above expressions can be combined and written compactly as:
\begin{eqnarray}
H({\bf k}) &= J& \left[ (2 \cos k_x + \cos k_y) \Gamma_1 +  \sin k_y \Gamma_2 \right. \nonumber \\ 
&& \left.+ (2 \cos k_z + \cos k_w) \Gamma_3 +  \sin k_w \Gamma_4\right] + M \Gamma_5 , \qquad \nonumber
\end{eqnarray} 
as stated in Eq.~\ref{eq:model} in the main text. 

Secondly, a similar procedure can be carried out to include the longer-range hopping terms given in the main text as
\begin{widetext}
\begin{eqnarray}
H_{\text{l}}& =&     J' \sum_{m,n,j,l} ( a^\dagger_{m+1,n+1,j+1,l+1} a_{m,n,j,l} +b^\dagger_{m+1,n+1,j+1,l+1} b_{m,n,j,l} -c^\dagger_{m+1,n+1,j+1,l+1} c_{m,n,j,l} -d^\dagger_{m+1,n+1,j+1,l+1} d_{m,n,j,l} + \rm{h.c} ) \nonumber \\
&&+ J'' \sum_{m,n,j,l} ( a^\dagger_{m+1,n+1,j-1,l-1} a_{m,n,j,l} +b^\dagger_{m+1,n+1,j-1,l-1} b_{m,n,j,l} -c^\dagger_{m+1,n+1,j-1,l-1} c_{m,n,j,l} -d^\dagger_{m+1,n+1,j-1,l-1} d_{m,n,j,l} + \rm{h.c} ) \nonumber
\end{eqnarray}
\end{widetext}
As above, we apply the Fourier transforms (Eq,~\ref{eq:fourier}); taking one term as an example, this gives:
\begin{eqnarray}
&& J'  \sum_{m,n,j,l} ( a^\dagger_{m+1,n+1,j+1,l+1} a_{m,n,j,l} +  \rm{h.c} ) \nonumber \\
&=& 2 J'  \sum_ {\bf k}  a^\dagger_{\bf k} \cos ( {\bf k} \cdot [ {\bf R}_1+  {\bf R}_2+  {\bf R}_3+ {\bf R}_4] )  a_{\bf k} \nonumber \\
&=& 2 J'  \sum_ {\bf k}  a^\dagger_{\bf k} \cos (2 k_x\!+ 2 k_z\! )  a_{\bf k}. \nonumber
\end{eqnarray}
where we have used the defined lattice vectors: 
${\bf R}_1 = (1, -1,0,0)$, ${\bf R}_2 = (1, 1,0,0)$, ${\bf R}_3 = (0, 0,1,-1)$ ${\bf R}_4 =(0, 0,1,1)$. 
Repeating this procedure for all the above long-range hopping terms, we arrive at a long-range-hopping momentum-space Hamiltonian which can be written compactly as
\begin{eqnarray}
H'({\bf k}) =   \left[ 2 J' \cos (2 k_x\!+ 2 k_z\! ) + 2 J'' \cos (2 k_x\!- 2 k_z\! )\right]  \Gamma_5, \qquad  \nonumber
\end{eqnarray}
as stated in Eq.~\ref{eq:mass} in the main text.

\end{document}